\documentstyle[twoside,12pt]{article}
\input{psfig}
\long\def\comment#1{}
\begin{document}
\title{Babylonian and Indian Astronomy: Early Connections}

\author{Subhash Kak\thanks{
Department of Electrical \& Computer Engineering,
Louisiana State University,
Baton Rouge, LA 70803-5901, USA,
Email: {\tt kak@ece.lsu.edu}}}

\date{February 17, 2003}
\maketitle

\section*{Introduction}

Did the Indian and Babylonian
astronomy evolve in
isolation, was there mutual influence, or was
one dependent on the other?
Scholars have debated these questions for
 more than two centuries, and opinion has
swung one way or the other with time.
The similarities between the two systems that
have been investigated are:
the use of 30 divisions of the lunar month;
the 360 divisions of the civil year;
the 360 divisions of the circle;
the length of the year; and the
solar zodiac.
Some have wondered if
the Babylonian planetary tables might have played a role 
in the theories of the siddh\={a}ntas.

I shall in this essay go over the essentials
of the early Indian and Babylonian astronomy and summarize the latest
views on the relationship between them.
I shall show that the key ideas found in the
Babylonian astronomy of 700 BC are already present 
in the Vedic texts, which even by the
most conservative reckoning are older than
that period.
I shall also show that the solar zodiac (r\={a}\'{s}is) was
used in Vedic India and I shall present a plausible derivation
of the symbols of the solar zodiac from the deities of
the segments.

In view of the attested presence of the
Indic people in the Mesopotamian region 
prior to 700 BC, it is likely that
if at all the two astronomies influenced
each other, the dependence is of
the Babylonian on the Indian. 
It is of course quite possible that the
Babylonian innovations emerged independent
of the earlier Indic methods. 

The Indic presence in West Asia goes back to the
second millennium BC in the ruling elites of the
Hittites and the Mitanni in Turkey and Syria,
and the Kassites in Mesopotamia.
The Mitanni were joined in marriage to the
Egyptian pharaohs during the second half
of the second millennium and they appear to
have influenced that region as well.$^{1}$
The Ugaritic list 33 gods, just like the count of
Vedic gods.

Although the Kassites vanished from the scene by
the close of the millennium, Indic
groups remained in the general area for centuries,
sustaining their culture by links through trade.
Thus Sargon defeats
one Bagdatti of Ui\v{s}di\v{s} in 716 BC. The name Bagdatti
(Skt. Bhagadatta) is Indic$^2$ and it cannot be Iranian because
of the double `t'.

The Indo-Aryan presence in West Asia persisted until
the time of the Persian Kings like Darius and Xerxes. It is
attested by the famous {\it daiva} inscription in which Xerxes
(ruled 486-465 BC)
proclaims his suppression of the rebellion by the daiva
worshipers of West Iran.

These Indic groups most likely served as intermediaries
for the transmission of ideas of Vedic astronomy to
the Babylonians and other groups in West Asia. 
Since we can clearly see a gap of several
centuries in the adoption of certain ideas, one can
determine the direction of transmission.
The starting point of
astronomical studies is the conception of
the wheel of time of 360 parts. It permeates
Vedic writing and belongs to 2nd millennium or the
3rd millennium BC or even earlier,
and we see it used in Babylon only in the second
part of first millennium BC.
Recent archaeological discoveries show that the
Sarasvat\={\i} river ceased reaching
the sea before 3000 BC and dried up in the sands
of the Western desert around 1900 BC, but this river is
praised as going from the mountain to the sea in
the \d{R}gveda.
This is consistent with astronomical evidence indicating
3rd millennium epoch for the \d{R}gveda.

\section*{Western Histories of Indian Astronomy}

The early Western studies of
Indian texts duly
noted the astronomical references to early epochs going
back to three or four thousand BC.
As the Indian astronomical texts were studied
it was discovered that the Indian methods were
different from those used in other civilizations.
The French astronomer M. Jean Sylvain Bailly 
in his classic {\it Trait\'{e} de l'Astronomie
Indienne et Orientale} (1787) described the methods of the
S\={u}rya Siddh\={a}nta and other texts and
expressed his view that Indian
astronomy was very ancient.
Struck by the elegance and
simplicity of its rules and its archaic features,
Bailly believed that astronomy had
originated in India and it was later transmitted to the Chaldeans
in Babylon and to the Greeks.

As against this, John Bentley in 1799 in a study in the
{\it Asiatick Researches} suggested that the parameters of
the S\={u}rya Siddh\={a}nta were correct for 1091 AD.
But Bentley was criticized for failing to notice that
the S\={u}rya Siddh\={a}nta had been revised 
using b\={\i}ja corrections,$^3$ and therefore his
arguments did not negate the central thesis of Bailly.

Meanwhile, in the next several decades
Indian astronomy became
a contested subject.
Part of the difficulty arose from a misunderstanding
of the Indian system due to the unfamiliar structure
of its luni-solar system.
Later, it became a hostage to the ideas
that the Vedic people had come as invaders to India around
1500 BC,$^4$ 
and that Indians were
otherworldly and uninterested in science and they
lacked the
tradition
of observational astronomy until the medieval times.$^5$
The inconvenient astronomical dates were brushed
aside 
as untrustworthy.
It was argued that astronomical references in the texts
either belonged to recent undatable layers or were late interpolations.$^6$

As against this, Ebenezer Burgess, the translator of the
S\={u}rya Siddh\={a}nta, writing in 1860, maintained that
the evidence, although not conclusive, pointed to the Indians being the 
original inventors or discoverers of:$^7$ (i) the lunar and solar
divisions of the zodiac, (ii) the primitive theory of epicycles,
(iii) astrology, and (iv) names of the planets after gods.$^8$

With the decipherment of the Babylonian astronomical tablets,
it was thought that early Indian astronomy may represent
lost Babylonian or Greek inspired systems.$^9$
But this leads to many difficulties, 
anticipated more than a hundred years earlier
by Burgess, including
the {\it incongruity} of the epochs involved.
This only thing that one can do is to lump all the
Indian texts that are prior to 500 BC together into a mass of uniform
material, as has been proposed by Pingree.$^{10}$
But such a theory is considered absurd by Vedic scholars.

The Ved\={a}\.{n}ga Jyoti\d{s}a, a late
Vedic text, belongs to the second millennium BC.
Although \'{S}a\.{n}kara B\={a}lak\d{s}\d{n}a D\={\i}k\d{s}ita's
{\it Bh\={a}rat\={\i}ya Jyoti\d{s}a},$^{11}$ published
in the closing years of the 19th century,
contained enough arguments against looking for any
foreign basis to the Ved\={a}\.{n}ga  Jyoti\d{s}a, 
the issue was reopened in the
1960s.$^{12}$
The basis behind rearticulation of an already
disproven theory was the idea that
``the origin of
mathematical astronomy in India [is] just one
element in a general transmission of Mesopotamian-Iranian
cultural forms to northern India during the two centuries
that
antedated Alexander's conquest of the Achaemenid empire.''$^{13}$

Overwhelming evidence
has since been furnished that disproves this
theory,$^{14}$ 
but many people remain confused about
the relationship between the two astronomy traditions.
The idea that India did not have a tradition
of observational astronomy
was refuted convincingly
by Roger Billard
more than thirty years ago. 
In his book on Indian astronomy,$^{15}$
he showed that the parameters used in the
various siddh\={a}ntas actually belonged to
the period at which they were created giving lie
to the notion that they were based on some old tables
transmitted from Mesopotamia or Greece.
The distinguished historian of astronomy
B.L. van der Waerden reviewed the ensuing controversy 
in a 1980 paper titled {\it Two treatises on Indian
astronomy} where he evaluated the views of Billard and
his opponent Pingree. He
ruled thus:$^{16}$
\begin{quote}
Billard's methods are sound, and his results shed new
light on the chronology of Indian astronomical treatises
and the accuracy of the underlying observations. We have
also seen that Pingree's chronology is wrong in several cases.
In one case, his error amounts to 500 years.
\end{quote}

For the pre-Siddh\={a}ntic period, the discovery
of the astronomy of the \d{R}gveda$^{17}$ establishes
that the Indians were making careful observations
in the Vedic period as well.

One might ask why should one even bother to revisit
Pingree's thesis if it stands discredited.
The reason to do so is that it provides a good context
to compare Babylonian and Indian astronomy and examine their
similarities and
differences. It also provides a lesson in how bad method
will lead to incongruous conclusions.

It is not my intention to replace Babylon by India as 
the source of
astronomical knowledge. I believe
that the idea of development in isolation is
simplistic; there existed much
interaction between the ancient
civilizations.
I also believe that
the borrowings in the ancient world were at best of
the general notions and the details of the astronomical
system that arose had features which made each system unique.
Rather than assign 
innovation to any specific group, we can at best 
speak of specific geographical areas in which, due to a variety of
social, economic, and cultural reasons, some new ways
of looking at the universe arose.
Regarding the problem of astronomy, we cannot ignore the
pre-Babylonian Indian literature
just as we must not ignore the fact that in the mid-first
millennium BC
the Babylonians
embarked on a notable period of careful astronomical records
and the use of mathematical models.$^{18}$
The Babylonian astronomical tradition goes back to the second
millennium BC or earlier but here we are concerned not with its remarkable
record of careful observation but with the beginnings of
mathematical astronomy which is ascribed to the middle first
millennium.

The next section will introduce pre-Ved\={a}\.{n}ga  Jyoti\d{s}a Indian astronomy
which will be followed by an account
of Babylonian astronomy so that the question of the
relationship between Ved\={a}\.{n}ga  Jyoti\d{s}a and Babylonian astronomy
may be investigated properly. Since the pre-Ved\={a}\.{n}ga  material belongs
mainly to the Sa\d{m}hit\={a}s that are squarely in the second
millennium BC or earlier epochs, it could not have been
influenced by Babylonian astronomy. We will also use the
evidence from the Br\={a}hma\d{n}as which also antedate the
Babylonian material in the most conservative chronology.

Once we have understood the nature of this earlier astronomy, we
will relate it to the 
Ved\={a}\.{n}ga  Jyoti\d{s}a and the Babylonian astronomies.

\section*{Pre-Ved\={a}\.{n}ga  Jyoti\d{s}a Astronomy}

Pre-Ved\={a}\.{n}ga  Jyoti\d{s}a astronomy was described at some length in my
essay titled ``Astronomy and its role in Vedic culture'' in volume 1
of the book$^{19}$ where the ritual basis of this science were sketched.
It was shown that the organization of the Vedic texts and the
altar ritual coded certain astronomical facts about the lunar and solar
years. This showed that observational astronomy was a part of
the tradition during the Vedic period itself.
But we will not invoke this knowledge here and restrict ourselves to
explicit statements from the Sa\d{m}hit\={a}s and the
Br\={a}hma\d{n}a literature.

The facts that emerge from the pre-Ved\={a}\.{n}ga  material include:
knowledge of the duration of the year, concept of tithi,
naming of ecliptic segments after gods, knowledge of
solstices for ritual, the 27- and 12- segment divisions
of the ecliptic, and the motions of the sun and the moon.

There were several traditions within the Vedic system.
For example, the month was reckoned in one with the
new moon, in another with the full moon.

\subsection*{Nak\d{s}atras}

Nak\d{s}atras stand for stars, asterisms or segments of the ecliptic.
The moon is conjoined with the 27 
nak\d{s}atras on successive nights in its passage around the earth; 
the actual cycle is of $27\frac{1}{3}$ days.
Because of this extra one-third day, there is drift in
the conjunctions that get corrected in three circuits. Also, 
the fact that the lunar year is shorter than the solar year
by 11+ days implies a further drift through the nak\d{s}atras
that is corrected by the use of intercalary months.

The earliest lists of
nak\d{s}atras in the Vedic books begin with K\d{r}ttik\={a}s, the Pleiades;
much later lists dating from sixth century AD begin with A\'{s}vin\={\i}
when the vernal equinox occurred on the
border of Revat\={\i} and A\'{s}vin\={\i}.
Assuming that the beginning of the list marked the same astronomical
event, as is supported by other evidence, 
the earliest lists should belong to the third
millennium BC or earlier. Each nak\d{s}atra has a presiding
deity (Taittir\={\i}ya Sa\d{m}hit\={a} 4.4.10).  In the Ved\={a}\.{n}ga Jyoti\d{s}a, the names of the
nak\d{s}atra and the deity are used interchangeably. 
It seems reasonable to assume that such usage had sanction 
of the tradition.

Table 1 provides a list of the nak\d{s}atras, the presiding deities, and
the approximate epoch for the winter and summer solstice for a few selected
nak\d{s}atras that are relevant to this paper. It is noteworthy that
the earliest Vedic texts provide us statements that recognize the 
movement of the solstices into new nak\d{s}atras. This provides us
a means to find approximate dates for these texts.
Our identification of the nak\d{s}atras has improved thanks
to the work of Narahari Achar$^{20}$ who has 
used powerful new simulation software
for sky maps that allows us to see the stars and the
planets in the sky as the Vedic people saw them.
Using this tool he has shown that some previous
identifications made without a
proper allowance for the shift in the ecliptic due to precession be modified.

The nak\d{s}atras in the Ved\={a}\.{n}ga Jyoti\d{s}a
represent 27 {\it equal parts} of the ecliptic. 
This appears to have been an old tradition since
the Sa\d{m}hit\={a}s (K\={a}\d{t}haka and Taittir\={\i}ya) mention
explicitly that Soma is wedded to all the nak\d{s}atras and
spends equal time with each.
The stars of
the nak\d{s}atras are thus just a guide to determine the
division of the ecliptic into equal parts. Each nak\d{s}atra
corresponds to 13$\frac{1}{3}$ degrees.

The following is a list of the nak\d{s}atras and their locations:

\begin{enumerate}
\item {\bf K\d{r}ttik\={a}}, from the root {\it k\d{r}t}, `to cut.'
These are the Pleiades, $\eta$ Tauri. 
{\it Deity:} Agni

\item {\bf Rohi\d{n}\={\i}}, `ruddy,' is $\alpha$ Tauri, Aldebaran.
{\it Deity:} Praj\={a}pati

\item {\bf M\d{r}ga\'{s}\={\i}r\d{s}a}, `Deer's head,' $\beta$ Tauri.
{\it Deity:} Soma 

\item {\bf \={A}rdr\={a}}, `moist,' is $\gamma$ Geminorum.
(Previously it was thought to be
Betelgeuse, $\alpha$ Orionis.)
{\it Deity:} Rudra

\item {\bf Punarvas\={u}}, `who give wealth again,' is the
star Pollux, or $\beta$ Geminorum.
{\it Deity:} Aditi

\item {\bf Ti\d{s}ya}, `pleased,' or {\bf Pu\d{s}ya}, `flowered,'
refers to $\delta$
Cancri in the middle of the other stars of this constellation.
{\it Deity:} B\d{r}haspati

\item {\bf \={A}\'{s}re\d{s}\={a}} or {\bf \={A}\'{s}le\d{s}\={a}},
`embracer,' represents $\delta, \epsilon, \zeta$ Hydrae.
{\it Deity:} Sarp\={a}\d{h}

\item {\bf Magh\={a}}, `the bounties,' is the group of stars near
Regulus, or $\alpha, \eta, \gamma, \zeta, \mu, \epsilon$ Leonis.
{\it Deity:} Pitara\d{h}

\item {\bf P\={u}rv\={a} Ph\={a}lgun\={\i}},
`bright,' $\delta$ and $\theta$ Leonis.
{\it Deity:} Aryaman (Bhaga)

\item {\bf Uttar\={a} Ph\={a}lgun\={\i}}, `bright,' $\beta$ and 93 Leonis.
{\it Deity:} Bhaga (Aryaman)

\item {\bf Hasta}, `hand.' The correct
identification is $\gamma$ Virginis. ( Previously,
the stars $\delta,~\gamma,~ \epsilon,~\alpha,~\beta$
in Corvus were assumed, but they are very far from the ecliptic
and thus not correctly located for this nak\d{s}atra.) 
{\it Deity:} Savitar

\item {\bf Citr\={a}}, `bright.' This is Spica or  $\alpha$ Virginis.
{\it Deity:} Indra (Tva\d{s}\d{t}\d{r})

\item {\bf Sv\={a}t\={\i}},
`self-bound,' 
 or {\bf Ni\d{s}\d{t}y\={a}},
is $\pi$ Hydrae. (The previous 
identification of Arctutus or $\alpha$ Bootis is too far from
the ecliptic.)
{\it Deity:} V\={a}yu

\item {\bf Vi\'{s}\={a}kh\={a}}, `without branches.' The stars $\alpha_2,
~\beta, \sigma$ Librae. 
{\it Deity:} Indr\={a}gni

\item {\bf Anur\={a}dh\={a}}, `propitious,' `what follows R\={a}dh\={a}.'
These are the $\beta, \delta, \pi$ Scorpii.
{\it Deity:} Mitra

\item {\bf Rohi\d{n}\={\i}}, `ruddy', or {\bf Jye\d{s}\d{t}h\={a}},
`eldest.' This is Antares, $\alpha$ Scorpii.
{\it Deity:} Indra (Varu\d{n}a)

\item {\bf Vic\d{r}tau}, `the two releasers,' or {\bf M\={u}la}, 
`root.' These are the stars from $\epsilon$ to $\lambda, \nu$ Scorpii.
{\it Deity:} Pitara\d{h} (Nir\d{r}ti)

\item {\bf P\={u}rv\={a} \={A}\d{s}\={a}\d{d}h\={a}},
`unconquered,' $\delta, \epsilon$ Sagittarii.
{\it Deity:} \={A}pa\d{h}

\item {\bf Uttar\={a} \={A}\d{s}\={a}\d{d}h\={a}},
`unconquered,' $\sigma, \zeta$ Sagittarii.
{\it Deity:} Vi\'{s}ve deva\d{h}

\noindent
{\bf Abhijit}, `reaching victory.'
The name refers to a satisfactory completion of the system
of nak\d{s}atras. The star is Vega, the brilliant $\alpha$ Lyrae.
This is the star that does not occur in the lists which have only
27 nak\d{s}atras on it.
{\it Deity:} Brahm\={a}

\item {\bf \'{S}ro\d{n}\={a}}, `lame,' or {\bf \'{S}rava\d{n}a}, `ear,'
$\beta$ Capricornus. (This is in place of
Altair, $\alpha$ Aquillae.)
{\it Deity:} Vi\d{s}\d{n}u

\item {\bf \'{S}ravi\d{s}\d{t}h\={a}},
`most famous.' Achar argues that it should be
$\delta$ Capricornus rather than the previously thought
$\beta$  Delphini.
It was later called {\bf Dhani\d{s}\d{t}h\={a}}, `most wealthy.'
{\it Deity:} Vasava\d{h}

\item {\bf \'{S}atabhi\d{s}aj}, `having a hundred physicians' is
$\lambda$ Aquarii and the stars around it.
{\it Deity:} Indra (Varu\d{n}a)

\item {\bf Pro\d{s}\d{t}hapad\={a}}, `feet of stool,' are the stars near
$\alpha$ Pegasi.
{\it Deity:} Aja Ekap\={a}d

\item {\bf Uttare Pro\d{s}\d{t}hapad\={a}}, `feet of stool,' and later
{\bf Bhadrapad\={a}}, `auspicious feet.'
These are $\gamma$ Pegasi and other nearby stars.
{\it Deity:} Ahirbudhnya

\item {\bf Revat\={\i}}, `wealthy,' $\eta$ Piscium.
{\it Deity:} P\={u}\d{s}an

\item {\bf A\'{s}vayujau}, `the two horse-harnessers,' are the stars
$\beta$ and $\alpha$ Arietis. {\bf A\'{s}vin\={\i}} is a later name.
{\it Deity:} A\'{s}vinau

\item {\bf Apabhara\d{n}\={\i}}, `the bearers,' are the group around
$\delta$
Arietis.
{\it Deity:} Yama

\end{enumerate}

The antiquity of the nak\d{s}atra system becomes clear when
it is recognized that all the deity names occur in RV 5.51 (this
insight is due to Narahari Achar$^{21}$). This hymn by Svasty\={a}treya
\={A}treya lists the deity names as:

\begin{quote}

A\'{s}vin, Bhaga, Aditi, P\={u}\d{s}an, V\={a}yu, Soma, 
B\d{r}haspati, SARVAGA\d{N}A\d{H}, 
Vi\'{s}ve Deva\d{h}, Agni, Rudra,
Mitra, Varu\d{n}a, Indr\={a}gni. 
\end{quote}

\vspace{0.2in}

The
sarvaga\d{n}a\d{h} are the ga\d{n}a\d{h} (groups)
such as the Vasava\d{h}, Pitara\d{h}, Sarpa\d{h} (including
Ahi and Aja), \={A}pa\d{h},
and 
the \={A}dityaga\d{n}a\d{h} (Dak\d{s}a Praj\={a}pati,
Aryaman, Vi\d{s}\d{n}u, Yama, Indra) complete
the list. There is no doubt that the
ecliptic is meant because the last verse of the
hymn refers explicitly to the fidelity with which
the sun and the moon move on their path,
the ecliptic.

The division of the circle into 360 parts or 720 parts was also
viewed from the point of view the nak\d{s}atras by
assigning 27 upanak\d{s}atras to each nak\d{s}atra
(\'{S}atapatha Br. 10.5.4.5).
This constituted an excellent approximation because
$27 \times 27 = 729$.
In other words, imagining each nak\d{s}atra to be further
divided into 27 equal parts made it possible to 
conceptualize half a degree when examining the sky.

The identification of the nak\d{s}atras is 
in consistent with their division into the two
classes of {\it deva} and {\it yama} nak\d{s}atras
as in the Taittir\={\i}ya Br\={a}hma\d{n}a 1.5.2.7:

\begin{quote} 

{\it k\d{r}ttik\={a}\d{h} prathama\d{m} vi\'{s}\={a}khe uttama\d{m}
t\={a}ni devanak\d{s}atr\={a}\d{n}i\\
anur\={a}dh\={a}\d{h} prathama\d{m}
apabhara\d{n}ihyuttama\d{m} t\={a}ni yamanak\d{s}atr\={a}\d{n}i\\
y\={a}ni devanak\d{s}atr\={a}\d{n}i t\={a}ni dak\d{s}i\d{n}ena pariyanti\\
y\={a}ni yamanak\d{s}atr\={a}\d{n}i t\={a}nyuttar\={a}\d{n}i iti.}\\

K\d{r}ttik\={a}s are the first and  Vi\'{s}\={a}khe are the last; those
are  {\it deva} nak\d{s}atras. Anur\={a}dh\={a}s are the first
and Apabhara\d{n}\={\i} is the last; those are the {\it yama} nak\d{s}atras.
The deva nak\d{s}atras revolve from the south; the yama nak\d{s}atras
revolve from the north.
\end{quote}

K\d{r}ttik\={a}s to Vi\'{s}\={a}khe are the
{\it deva} nak\d{s}atras because they lie north of
the equator, whereas the others are {\it yama}
nak\d{s}atras because they lie south of the equator.
Since the devas are supposed to live in the north pole and
Yama in the south pole, the deva nak\d{s}atras revolve south
of the abode of the devas, and the yama nak\d{s}atras revolve
north of the abode of Yama.
This classification helps confirm the identification of the
nak\d{s}atras.

\vspace{0.3in}
\noindent
{\it Table 1:} Nak\d{s}atras with their Deity names and the approximate 
epoch of winter solstice and spring equinox at the midpoint of each segment

\vspace{0.3in}

\begin{tabular}{||r|l|l|l|l||}\hline
{\it Num} & {\it Nak\d{s}atra} & {\it Deity} & {\it W. Solstice}& {\it S. Equinox}\\ \hline
1 & K\d{r}ttik\={a} & Agni &  & 2000 BC \\

2 & Rohi\d{n}\={\i} & Praj\={a}pati& & 3000 BC \\

3 & M\d{r}ga\'{s}\={\i}r\d{s}a & Soma & & 4000 BC \\

4 & \={A}rdr\={a} & Rudra & & 5000 BC \\

5 & Punarvas\={u} & Aditi & & 6000 BC \\

6 & Ti\d{s}ya or Pu\d{s}ya & B\d{r}haspati & & \\

7 & \={A}\'{s}re\d{s}\={a} or  \={A}\'{s}le\d{s}\={a} & Sarp\={a}\d{h} & &\\

8 & Magh\={a} & Pitara\d{h}& & \\

9 & P\={u}rv\={a} Ph\={a}lgun\={\i} & Aryaman & &\\

10 & Uttar\={a} Ph\={a}lgun\={\i} & Bhaga & &\\

11 & Hasta & Savitar & &\\

12 & Citr\={a} & Indra & &\\

13 & Sv\={a}t\={\i}  or Ni\d{s}\d{t}y\={a} & V\={a}yu & &\\

14 & Vi\'{s}\={a}kh\={a} & Indr\={a}gni & &\\

15 & Anur\={a}dh\={a} & Mitra & &\\

16 & Rohi\d{n}\={\i} & Indra & &\\

17 & Vic\d{r}tau or M\={u}la & Pitara\d{h}   & 2000 AD &\\

18 & P\={u}rv\={a} \={A}\d{s}\={a}\d{d}h\={a} & \={A}pa\d{h}  & 1000 AD &\\

19 & Uttar\={a} \={A}\d{s}\={a}\d{d}h\={a} & Vi\'{s}ve deva\d{h}& 0 AD &\\

* &  Abhijit & Brahm\={a} & & \\

20 & \'{S}ro\d{n}\={a} or  \'{S}rava\d{n}a & Vi\d{s}\d{n}u &1000 BC &\\

21 & \'{S}ravi\d{s}\d{t}h\={a} or Dhani\d{s}\d{t}h\={a} & Vasava\d{h} & 2000 BC &\\

22 & \'{S}atabhi\d{s}aj & Indra  & 3000 BC &\\

23 & Pro\d{s}\d{t}hapad\={a} & Aja Ekap\={a}d & 4000 BC &\\

24 & Uttare Pro\d{s}\d{t}hapad\={a} & Ahirbudhnya  & 5000 BC & 2000 AD\\

25 & Revat\={\i} & P\={u}\d{s}an  & 6000 BC & 1000 AD\\

26 & A\'{s}vayujau & A\'{s}vinau & 7000 BC & 0 AD\\

27 & Apabhara\d{n}\={\i} & Yama & & 1000 BC \\ \hline

\end{tabular}

\newpage

Abhijit, which comes between
the nineteenth and the twentieth in the above list, 
does not occur in the list of
the 27 in Taittir\={\i}ya Sa\d{m}hit\={a}
or in Ved\={a}\.{n}ga Jyoti\d{s}a. Maitr\={a}y\={a}\d{n}\={\i}
and K\={a}\d{t}haka Sa\d{m}hit\={a}s and Atharvaveda contain
lists with the 28 nak\d{s}atras.

When the asterisms K\d{r}ttik\={a} and Vi\'{s}\={a}kh\={a} defined
the spring and the autumn equinoxes, the asterisms Magh\={a} and
\'{S}ravi\d{s}\d{t}h\={a} defined the summer and the winter
solstices.

\subsection*{The Year and Solstices}

There were two kinds of year in use. In one, the year was measured
from one winter solstice to another; in the other, it was measured from
one vernal equinox to another. Obviously, these years were solar
and related to the seasons (tropical).

The wheel of time was defined to have a period of 360 parts. This number seems
to have been chosen as the average of 354 days of the lunar year and
the 366 days for the solar year.

In TS 6.5.3, it is said that the sun travels moves northward
for six months and southward for six months. 
The Br\={a}hma\d{n}as speak of ritual that follows the course of the
year starting with the winter solstice.
For example, the Pa\~{n}cavi\d{m}\'{s}a Br\={a}hma\d{n}a
describes sattras of periods of several days, as well as
one year (PB 25.1), 12 years, 1000 days, and 100 years.
In these types of ritual the number of days were recorded, 
providing a means of determining an accurate size of the
solar year.
The sattra of 100 years appears to refer to the centennial system of
the Saptar\d{s}i calendar.

The solstice day was probably determined by the noon-shadow of
a vertical pole. The Aitareya Brahmana speaks of the sun remaining
stationary for about 21 days at its furthest point in the north (summer
solstice) and likewise for its furthest point in the south (winter
solstice). 
This indicates that the motion of the sun was not taken to be
uniform all the time.

\subsection*{Months}
The year was divided into 12 months which were defined with respect
to the nak\d{s}atras, and with respect to the movements of the moon.

The Taittir\={\i}ya Sa\d{m}hit\={a} (TS) (4.4.11) gives a list of solar
months:

\begin{quote}
Madhu, M\={a}dhava (Vasanta, Spring), \'{S}ukra, \'{S}uci (Gr\={\i}\d{s}ma,
Summer), Nabha, Nabhasya (Var\d{s}\={a}, Rains), I\d{s}a and \={U}rja 
(\'{S}arad, autumn),
Sahas and Sahasya (Hemanta, Winter), and Tapa and Tapasya (\'{S}i\'{s}ir, Deep
Winter).
\end{quote}

The listing of months by the season implies that parts of
the ecliptic were associated with these 12 months.  These
months are also known by their
\={A}ditya names (Table 2). These names vary from text to text,
therefore, we are speaking of more than one tradition.
It should be noted that different lists of names need not
mean usage at different times.

\vspace{0.2in}
\noindent
{\it Table 2:} The twelve months with the nak\d{s}atra
named after and \={A}dityas  names
(from Vi\d{s}\d{n}u Pur\={a}\d{n}a)

\vspace{0.3in}

\begin{tabular}{||l|l|l||}\hline
{\it Month} & {\it Nak\d{s}atra}  & {\it \={A}ditya}\\ \hline
Caitra &  Citr\={a}  & Vi\d{s}\d{n}u \\
Vai\'{s}\={a}kha & Vi\'{s}\={a}kh\={a} &  Aryaman \\
Jyai\d{s}\d{t}ha & Jye\d{s}\d{t}h\={a}  & Vivasvant\\
\={A}\'{s}\={a}\d{d}ha & \={A}\'{s}\={a}\d{d}h\={a}s  & A\d{m}\'{s}u \\
\'{S}r\={a}va\d{n}a & \'{S}ro\d{n}a  & Parjanya\\
Bh\={a}drapada & Pro\d{s}\d{t}hapadas  & Varu\d{n}a \\
\={A}\'{s}vayuja & A\'{s}vin\={\i}  & Indra\\
K\={a}rtika &K\d{r}ttik\={a}  & Dh\={a}t\d{r}\\
M\={a}rga\'{s}\={\i}r\d{s}a & M\d{r}ga\'{s}iras  & Mitra\\
Pau\d{s}a & Ti\d{s}ya  & P\={u}\d{s}an\\
M\={a}gha & Magh\={a}  & Bhaga\\
Ph\={a}lguna & Ph\={a}lgun\={i}  & Tva\d{s}\d{t}\={a}\\ \hline

\end{tabular}

\vspace{0.3in}

Now we investigate if the r\={a}\'{s}i names associated with
the segments were a part of the Vedic tradition or if they
were adopted later. In any adoption from Babylonia or Greece,
one would not expect a fundamental continuity with the
nak\d{s}atra system.
Taking the clue from the Ved\={a}\.{n}ga Jyoti\d{s}a, where
the names of the nak\d{s}atras and the deities are used
interchangeably, we will investigate if the r\={a}\'{s}i
names are associated with the segment deities.

The nak\d{s}atra names of the months each cover 30$^o$ of
the arc, as against the 13$\frac{1}{3} ^o$ of the lunar
nak\d{s}atra segment.
Therefore, the extension of each month may stretch over upto
three
nak\d{s}atras with corresponding deities. 
This will be seen in Figure 1 or in the list below.
The choice made in Figure 1, where Vai\'{s}\={a}kha
begins with the 
the sun in the ending segment of A\'{s}vin\={\i} and
the moon at the mid-point of Sv\={a}t\={\i}
is the most likely assignment as it bunches the
\={A}\'{s}\={a}\d{d}h\={a}s and the Ph\={a}lgun\={\i}s
in the right months, with the Pro\d{s}\d{t}hapad\={a}s
three-fourths correct and \'{S}ro\d{n}\={a} half-correct.
The full-moon day of the lunar month will thus fall
into the correct nak\d{s}atra.
Since the solar and the
lunar months are not in synchrony, the mapping would tend to slip upto
two nak\d{s}atra signs
until it is corrected by the use of the intercalary
month.
At worst, we get
a sequence of r\={a}\'{s}is which is out of step by one.

\begin{figure}
\hspace*{0.0in}\centering{
\psfig{file=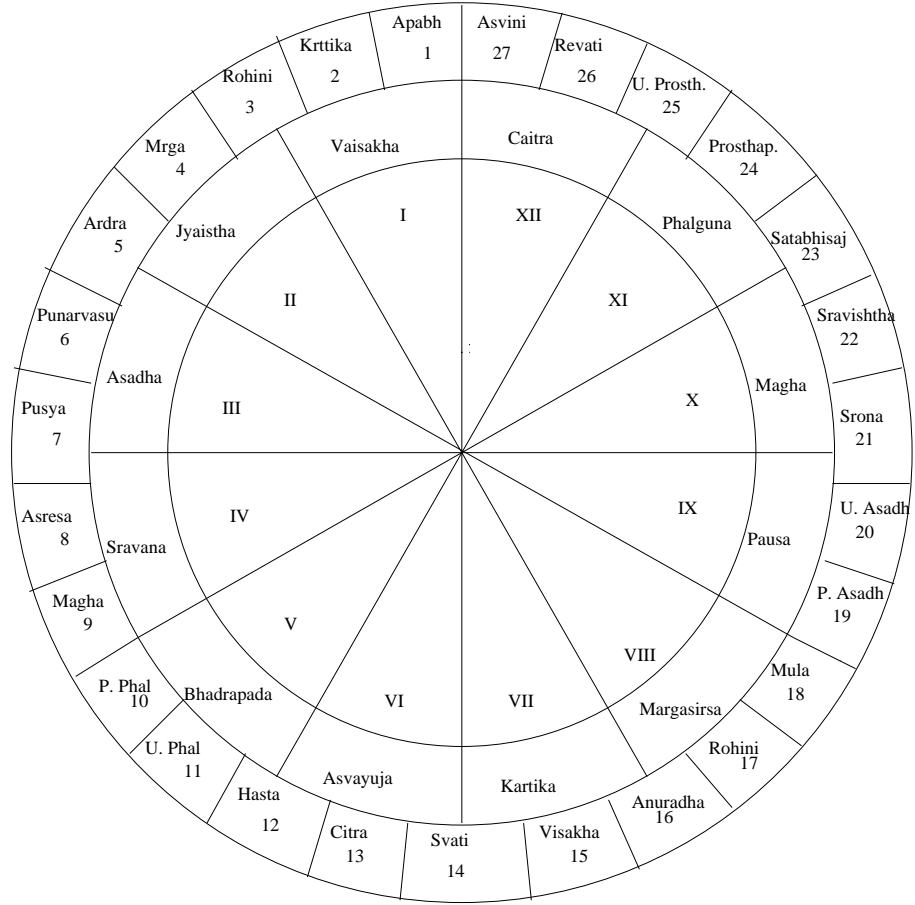,width=12cm}}
\caption{The 27-fold and 12-fold division of the ecliptic. The first
r\={a}\'{s}i is V\d{r}\d{s}a with the corresponding month of Vai\'{s}\={a}kha}
\end{figure}

\vspace{0.3in}

\begin{quote}

Vai\'{s}\={a}kha = Sv\={a}t\={\i} to Anur\={a}dh\={a} = V\={a}yu, Indr\={a}gni,
Mitra\\ = 
V\d{r}\d{s}a, Bull for Indra, e.g. RV 8.33; also V\={a}yu is sometimes
identified with Indra and the two together called Indrav\={a}y\={u}, and
V\={a}yu is also associated with cow (RV 1.134)\\

Jyai\d{s}\d{t}ha = Anur\={a}dh\={a} to M\={u}la = Mitra, Varu\d{n}a, Pitara\d{h} \\
=  Mithuna, Gemini, from the cosmic embrace of Mitra and Varu\d{n}a \\

\={A}\'{s}\={a}\d{d}ha = P\={u}rva \={A}\'{s}\={a}\d{d}h\={a}
to \'{S}ro\d{n}a =
\={A}pa\d{h}, Vi\'{s}ve Deva\d{h}, Vi\d{s}\d{n}u \\
=  Karka, circle or Cancer, the sign of Vi\d{s}\d{n}u's cakra
(e.g. RV 1.155.6)\\

\'{S}r\={a}va\d{n}a =
\'{S}ro\d{n}a to \'{S}atabhi\d{s}aj =
Vi\d{s}\d{n}u, Vasava\d{h}, Indra \\= 
Si\d{m}ha, Lion, after
Indra as in RV 4.16.14\\

Bh\={a}drapada =
\'{S}atabhi\d{s}aj to U.
 Pro\d{s}\d{t}hapada =
Indra, Aja Ekap\={a}da, Ahirbudhnya \\=  
Kany\={a}, Virgin, apparently from Aryaman in the opposite side
of the zodiac who is the wooer of maidens, kany\={a} (RV 5.3.2)\\

\={A}\'{s}vina =
 U. Pro\d{s}\d{t}hapada to A\'{s}vayujau =
Ahirbudhnya, P\={u}\d{s}an, A\'{s}vayujau \\= 
Tul\={a}, Libra, from the \={A}\'{s}vins who denote
balance of pairs (e.g. RV 2.39, 5.78, 8.35) \\

K\={a}rtika =
Apabhara\d{n}\={\i} to Rohi\d{n}\={\i} =
Yama, Agni, Praj\={a}pati \\=
Ali (V\d{r}\'{s}cika), Scorpion, from K\d{r}ttika, to cut\\

M\={a}rga\'{s}\={\i}r\d{s}a =
Rohi\d{n}\={\i} to \={A}rdr\={a} = Praj\={a}pati, Soma, Rudra \\= 
Dhanu\d{s}, Archer, from the cosmic archer Rudra (RV 2.33, 5.42, 10.125)\\

Pau\d{s}a =
\={A}rdr\={a} to Pu\d{s}ya = Rudra, Aditi, B\d{r}haspati \\= 
Makara, Goat, Rudra placing
goat-head on Praj\={a}pati, and goat is the main animal
sacrificed at the ritual of which B\d{r}haspati is the priest\\

M\={a}gha =
Pu\d{s}ya to Magh\={a} =
B\d{r}haspati, Sarpa\d{h}, Pitara\d{h} \\=
Kumbha, Water-bearer, from the water-pot offerings to the pitara\d{h} \\

Ph\={a}lguna =
Ph\={a}lgun\={\i}s  to Hast\={a} = Aryaman, Bhaga, Savitar \\= 
M\={\i}na, Fish, representing Bhaga (alluded to in RV 10.68)\\

Caitra = Hast\={a} to Sv\={a}t\={\i} = Savitar, Indra, V\={a}yu \\= 
Me\d{s}a, Ram, from Indra, see, e.g., RV 1.51\\

\end{quote}

\vspace{0.3in}

We observe that for most solar zodiac segments a plausible  name
emerges from the name of the deity.
The choice of the symbols was also governed by another constraint.
The Br\={a}hma\d{n}a texts call the year as the sacrifice and
associate different animals with it.$^{22}$
In the short sequence, these animals are goat, sheep, bull, horse, and man.
Beginning with the goat-dragon at number 9 in the sequence
starting with Vai\'{s}\={a}kha, we have sheep at 12,
bull at 1, horse (also another name for the
sun in India) as the sun-disk at 3, and man as archer at 8.

\subsection*{Intercalation}

A system of intercalation of months (adhikam\={a}sa) was used
to bring the lunar year in synchrony with the solar year over
a period of five years.

The use of the intercalary month (adhikam\={a}sa) goes back to 
the \d{R}gveda itself:
\begin{quote}
{\it vedam\={a}so dh\d{r}tavrato dv\={a}da\'{s}a praj\={a}vata\d{h}\\
ved\={a} ya upaj\={a}yate} (RV 1.25.8)\\

Dh\d{r}tavrata (Varu\d{n}a) knew the twelve productive months; he 
also knew about the thirteenth additional month.
\end{quote}

In the Atharvaveda (13.3.8), it is said:
\begin{quote}

{\it ahor\={a}traivimita\d{m} tri\d{m}\'{s}ada\.{n}ga\d{m}\\
trayoda\d{s}a\d{m} m\={a}sa\d{m} yo nirmim\={\i}te} (AV 13.3.8)\\

He who forms the thirteenth month containing thirty days and nights.
\end{quote}

The names of the two intercalary months are given as
{\it sa\d{m}sarpa} and {\it a\d{m}haspati} in the
Taittir\={\i}ya Sa\d{m}hit\={a} 1.4.14.

There are several other similar references in the Sa\d{m}hit\={a}
literature to the various intercalary schems that were used
to reconcile the lunar and solar years.

\subsubsection*{The concept of yuga}

The \d{R}gveda mentions yuga in what is most likely
a five-year sense in RV 1.158.6.
The names of two of these five years,
{\it sa\d{m}vatsara} and {it parivatsara} are to be found in
RV 7.103.7.
The V\={a}jasaneyi Sa\d{m}hit\={a} (27.45 and 30.16) 
and the Taittir\={\i}ya Sa\d{m}hit\={a} (5.5.7.1-3) give
the names of all the five years.
These names are: sa\d{m}vatsara, parivatsara,
id\={a}vatvara, iduvatsara, and vatsara.

The number five is fundamental to Vedic imagination. Thus there are
five-layers of the altar, five breaths within man, five seasons,
and five kinds of sacrifices.
It was natural then to conceive of a five-year yuga as a basic
period since larger yugas were known.

The use of the five year yuga is natural to do a basic
synchronization of the lunar and the solar years. Longer
periods are required for a more precise synchronization rules.

\subsection*{Circle of 360$^o$}

In \d{R}gveda 1.164.11, mention is made
of the 720 paired sons of the wheel of time which has twelve spokes.
These 720 pairs are the 720 days and nights of the civil year.
 In RV 1.164.48 we are explicitly told of
the 360 parts of the wheel of time.

\begin{quote}
 {\it dv\={a}da\'{s}a pradhaya\'{s} cakram eka\d{m}\\
 tr\={\i}\d{n}i nabhy\={a}ni ka utacciketa\\
 tasmin s\={a}ka\d{m} tri\'{s}at\={a} na \'{s}a\.{n}kavo\\
 arpit\={a}\d{h} \d{s}a\d{s}\d{t}irna cal\={a}cal\={a}sa\d{h}} (RV 1.164.48)\\

  Twelve spokes, one wheel, three navels, who can comprehend this?
In this there are 360 spokes put in like pegs which do not get loosened.
\end{quote}

This means that the ecliptic, which is the
wheel of time, is 
divided into 360 parts.
Each of these parts is what is
now known as a degree.
The three navels appear to be the three different kinds of divisions of it:
solar and lunar segments and days.

The division of the circle into four quadrants of 90 degrees each is
described in another hymn: 

\begin{quote}
{\it caturbhi\d{h} s\={a}k\d{m} navati\d{m} ca n\={a}mabhi\'{s} cakra\d{m} 
na v\d{r}tta\d{m} vyat\={i}\d{d}r av\={\i}vipat} (RV 1.155.6)\\

He, like a rounded wheel, hath in swift motion set his ninety racing steeds 
together with the four.
\end{quote}

The division of the wheel of time into 360 parts
occurs elsewhere as well. In \'{S}atapatha Br. 10.5.4.4,
it is stated that ``360 regions encircle the sun on all sides.''

The division into half a degree is very easy to identify in the
sky. The radial size of the sun or moon is slightly more
than this angular size, being exactly 60/113 degrees.$^{23}$

Note, further, that the day is divided into 60 n\={a}\d{d}ikas in
the Ved\={a}\.{n}ga Jyoti\d{s}a.
Since the day is to the year what the degree is to the circle, 
this means that the degree was further divided into 60 parts.

\subsection*{Various Divisions of the Ecliptic}
One may argue that because the original list of
27 nak\d{s}atras contains only 24 distinct names, these represent
the 24 half months of the year. Later, to
incorporate lunar conjunctions, 
the segments were expanded to describe the motions of the moon.

In the \d{R}gveda (2.27), six \={A}dityas  are listed which appear
to be segments corresponding to the
six seasons. The  names given are: Mitra, Aryaman, 
Bhaga, Varu\d{n}a, Dak\d{s}a, A\d{m}\'{s}a.

This notion is supported  by the fact that the
ecliptic is also described in terms of the twelve \={A}dityas as
in Table 3.
In the \'{S}atapatha Br\={a}hma\d{n}a (6.1.2.8),
Praj\={a}pati is said to have ``created the twelve \={A}dityas,
and placed them in the sky.''
In \'{S}atapatha Br. (11.6.3.8), it is stated that the \={A}dityas
are the twelve months ({\it dv\={a}da\'{s}a m\={a}sa\d{h}}).
This means clearly a twelve part division of the circuit of the sun.

The correspondence between the 27-fold division and the
12-fold division of the ecliptic may be seen in Figure 1.

Further division of the ecliptic is seen in the subdivision
of each of the r\={a}\'{s}is into
2, 3, 4, 7, 9,
10, 12, 16, 20, 24, 27, 30, 40, 45, 45, and 60 parts.

\subsection*{Nak\d{s}atras and chronology}
The list beginning with K\d{r}ttik\={a} at the
vernal equinox indicates
that it was drawn up in the third millennium BC.
The legend of the decapitation of Praj\={a}pati indicates
a time when the year began with M\d{r}ga\'{s}\={\i}r\d{s}a
in the fifth millennium BC (Table 1).
Scholars have also argued that a subsequent list began with
Rohi\d{n}\={\i}. This reasoning is supported by the fact that there are
two Rohi\d{n}\={\i}s, separated by fourteen nak\d{s}atras,
indicating that the two 
marked the beginning of the two half-years.

In addition to the chronological implications of the changes
in the beginning of the Nak\d{s}atra lists, there are other
references which indicate epochs that bring us down to the
Common Era.

The moon rises at the time of sunset on p\={u}r\d{n}im\={a},
the full moon day.
It rises about 50 minutes every night and at the end of
the \'{S}ivar\={a}tri of the month, about two
days before am\={a}vasy\={a}, it rises about an hour
before sunrise. The crescent moon appears first above the horizon,
followed by the rising sun. This looks like the sun as \'{S}iva
with the crescent moon adorning his head.
This is the last appearance of the moon in the month before its
reappearance on \'{s}ukla dvit\={\i}ya. These two days were
likely used to determine the day of am\={a}v\={a}sya.

Mah\={a}\'{s}ivar\={a}tri is the longest night of the year
at the winter solstice.
At present, this occurs on February 26 $\pm$ 15 days (this uncertainty
arises from the manner in which the
intercalary month operates), and when it was introduced
(assuming a calendar similar to the present one), the epoch would
have been December 22 $\pm$ 15 days. The difference of 66 days gives
an epoch of 2600 BC $\pm$ 1100 years for the establishment of
this festival.$^{24}$

The Kau\d{s}\={\i}taki Br. (19.3) mentions the occurrence of
the winter solstice in the new
moon of
M\={a}gha ({\it m\={a}ghasy\={a}m\={a}v\={a}sy\={a}y\={a}m}). 
This corresponds to a range of 1800---900 BC based on 
the uncertainty related to the precise identification of the Magh\={a}
nak\d{s}atra at that time.

The \'{S}atapatha Br\={a}hma\d{n}a (2.1.2.3) has a statement
that points to an earlier epoch where it is
stated that K\d{r}ttik\={a} never swerve from the
east.
This corresponds to 2950 BC.
The Maitray\={a}n\={\i}ya Br\={a}hma\d{n}a Upani\d{s}ad
(6.14) refers to the winter
solstice being at the mid-point of the
\'{S}ravi\d{s}\d{t}h\={a} segment and the
summer solstice at the beginning of Magh\={a}.
This indicates 1660 BC.

The Ved\={a}\.{n}ga Jyoti\d{s}a (Yajur 6-8) mentions that winter
solstice was at
the beginning of
\'{S}ravi\d{s}\d{t}h\={a}
and the summer solstice at the mid-point of
A\'{s}le\d{s}\={a}.  This corresponds to about 1350 BC
if the nak\d{s}atra is identified with 
$\beta$ Delphini and to 1800 BC if it is identified
with $\delta$ Capricornus, the more correct assignment.$^{25}$

In TS 7.4.8 it is stated that the year begins with the
full moon of Ph\={a}lgun\={\i}. In the Ved\={a}\.{n}ga 
Jyoti\d{s}a, it begins with the full moon in Magh\={a},
providing further evidence forming a consistent whole.

The \'{S}atapatha Br\={a}hma\d{n}a story of the marriage between the 
Seven Sages, the stars of the Ursa Major, and the K\d{r}ttik\={a}s
is elaborated in the Pur\={a}\d{n}as where it is stated that the
\d{r}\d{s}is remain for a hundred years in each nak\d{s}atra.
In other words, during the earliest times in India there existed a
centennial calendar with a cycle of 2,700 years.
Called the Saptar\d{s}i calendar, 
it is still in use in several parts of India. Its current beginning is taken to
be 3076 BC, but the notices by the Greek historians Pliny and
Arrian suggest that, during the 
Mauryan times, this calendar was taken to begin in 6676 BC.

\section*{Babylonian Astronomy}
Our knowledge of Babylonian astronomy comes from three
kinds of texts.
In the first class are: (i) astronomical omens in the style of
{\it En\={u}ma Anu Enlil} (``when the gods Anu and Enlil'')
that go back to
the second millennium BC in a series of 70 tablets;
(ii) the two younger Mul Apin tablets which is more astronomical;
(iii) royal reports on omens from 700 BC onwards.

The second class has astronomical diaries with excellent observations
over the period 750 BC to AD 75. The third class has texts
from the archives in Babylon and Uruk from the period of the
last four or five centuries BC which deal with mathematical
astronomy.

In late texts the ecliptic is divided into 12 zodiacal signs,
each of length precisely 30 degrees.
Aaboe has proposed$^{26}$ that the replacement of constellations
by 30$^o$ segments took place in the
fifth century BC.

Babylonian mathematics is sexagesimal, that is, it uses a
place-value system of base 60.
This is considered one of the characteristic features
of the Babylonian mathematical tradition.

The Babylonian year began with or after vernal equinox.
The calendar is lunar with a new month beginning
on the evening when the crescent of the new moon becomes
visible for the first time.
A month contains either 29 days (hollow)
or 30 days (full). Since 12 lunar months add up to only 354 days,
an intercalary month was occasionally introduced. Starting
mid-fifth century, the intercalations followed the Metonic cycle
where every group of 19 years contained seven years with 
intercalary months.

In the late texts the ecliptic is divided into 12 zodiacal
signs, each of length precisely 30 degrees (u\v{s}).
The first list of stars which used the signs of the zodiac
is dated to about 410 BC.

The zodiacal signs have much overlap with the Indian ones, but they
appear from nowhere. We cannot, for example, understand the basis of
goat-fish, whereas the goad-headed Praj\={a}pati is one of the
key stories in Vedic lore.
These signs do not belong to the same type. They
include furrow, 
hired hand, and star.
They could not have served as the model for the Indian zodiacal
names or the Greek ones because of their haphazard nature.
On the other hand, they could represent memory of an
imperfectly communicated Indian tradition which was adapted 
into the Babylonian system. 
The Indic kingdoms in West Asia in the second
millennium BC could have served as the intermediaries in such
transmission.

{\it Table 3}: The Zodiac signs

\vspace{0.3in}
\begin{tabular}{|l|l|l|} \hline
{\it Latin} & {\it Babylonian} & {\it Greek} \\ \hline \hline
Aries & hun, lu (hired hand) & Krios (ram) \\ \hline
Taurus & m\'{u}l (star) & Tauros (bull) \\ \hline
Gemini & mash, mash-mash (twins) & Didymoi (twins) \\ \hline
Cancer & alla$_x$, ku\v{s}u (?) & Karkinos (crab)\\ \hline
Leo & a (lion) & Leon (lion) \\ \hline
Virgo & absin (furrow) & Parthenos (virgin)\\ \hline
Libra & r\'{i}n (balance) & Khelai (claws) \\ \hline
Scorpio & g\'{i}r (scorpion) &  Skorpios (scorpion) \\ \hline
Sagittarius & pa (name of a god) & Toxotes (archer) \\ \hline
Capricornus & m\'{a}\v{s} (goat-fish) & Aigokeros (goat-horned) \\ \hline
Aquarius & gu (?) & Hydrokhoos (water-pourer) \\ \hline
Pisces & zib, zib-me (tails) & Ikhthyes (fishes)\\ \hline
\end{tabular}

\vspace{0.3in}

The Babylonians had two systems to place the signs on the ecliptic. In one, 
the summer solstice was at 8$^o$ in ku\v{s}u (and the
winter solstice in 8$^o$ in m\'{a}\v{s}); in another system, the
solstices were at 10$^o$ of their signs.
They measured the moon and the planets from the ecliptic using
a measure called {\it she}, equal to $1/72$ of a degree.

They appear to have used two models for the sun's motion. In one,
the sun's velocity changes suddenly; in another, it goes through a
zig-sag change.

As far as planets are concerned, they calculated the dates of the
instants the planet starts and ends its retrogression,
the first visible heliacal rising, 
the last visible heliacal rising, and opposion. They
also computed the position of the planet on the ecliptic
at these instants.
In the planetary theory, the synodic month is divided into
30 parts, which we now call {\it tithi} from its Indian usage.

In the Babylonian planetary models the concern is to 
compute the time and place of first stationary points.
Two different theories to do this were proposed which have
been reconstructed in recent decades.$^{27}$

\section*{Babylonian Astronomy and the Ved\={a}\.{n}ga Jyoti\d{s}a}

The thesis that Babylonian astronomy may have
led to Vedic astronomy was summarized in the following manner
by David Pingree:$^{28}$
\begin{quote}
Babylonian astronomers were capable of devising
intercalation-cycles in the seventh, sixth, and fifth
centuries B.C., and there is evidence both in the Greek
and in the cuneiform sources that they actually did so;
and by the early fourth century B.C. they had certainly
adopted the quite-accurate nineteen-year cycle.
It is my suggestion that some knowledge of these attempts
reached India, along with the specific astronomical
material in the fifth or fourth century B.C.
through Iranian intermediaries, whose influence is
probably discernible in the year-length selected by
Lagadha for the {\it Jyoti\d{s}aved\={a}\.{n}ga}.
But the actual length of the yuga, five years, was
presumably accepted by Lagadha because of its identity
with a Vedic lustrum. Not having access to a series
of extensive observations such as were available to
the Babylonians, he probably was not completely
aware of the crudeness of his system. And the
acceptance of this cycle by Indians for a period of
six or seven centuries or even more demonstrates among
other things that they were not interested in performing
the simplest acts of observational astronomy.
\end{quote}

The specific items from Babylonian astronomy that
Pingree believes were incorporated into the
``later'' Vedic astronomy are :

\begin{enumerate}

\item 
The ratio of 3:2 for the longest to the shortest day
used after 700 BC. 

\item
The use of a linear function to determine the
length of daylight in intermediate months.

\item
The use of the water-clock.

\item
The concept of the {\it tithi} as the
thirtieth part of the lunar month.

\item
The use of two intercalary months in a period of 5 years.

\item
The concept of a five-year yuga.

\end{enumerate}

Each of these points has been answered by several historians.
In particular, T.S. Kuppanna Sastry wrote a much-acclaimed
text on the Ved\={a}\.{n}ga Jyoti\d{s}a showing how the
supports its dating of around 1300-1200 BC, and now Achar
has argued for its dating to about 1800 BC.
In fact, in his classic {\it Bh\={a}rat\={\i}ya
Jyoti\d{s}a} (1896), S.B. D\={\i}k\d{s}ita had already
documented the Vedic roots of Vedic astronomy.
More recently, 
Achar$^{29}$ has dealt with these questions at length in
his paper on the Vedic origin of ancient mathematical
astronomy in India.

\subsubsection*{Length of the Day}

The proportion  of 3:2 for the longest to the shortest day
is correct for northwest India. On the other hand, the 
Babylonians until 700 BC or so used the incorrect
proportion of 2:1. It is clear then that the Babylonians
for a long time
used a parameter which was completely
incorrect. They must have, therefore, revised this parameter
under the impulse of some outside influence.

In any event, the 3:2  proportion proves nothing because it
is correct both for parts of India and Babylon. Its late usage
in Babylonia points to the limitations of Babylonian observational
astronomy before 700 BC.

\subsubsection*{The Use of a Linear Function for Length of Day}

The interpolation formula in the
\d{R}gjyoti\d{s}a, verse 7, is:


$d(x) = 12 + 2 x /61$

\noindent
where $d$ is the duration of day time in muh\={u}rtas and
$x$ is the number of days that have elapsed since the winter
solstice.

The use of this equation is natural when one considers the
fact that the number of muh\={u}rtas required for the
winter solstice for the 3:2 proportion to hold is 12.
This ensures that the length of day and night will be
equal to 15 muh\={u}rtas each at the equinox.

The Taittir\={\i}ya Sa\d{m}hit\={a} 6.5.3.4 speaks clearly
of the northern and southern movements of the sun:
{\it \={a}ditya\d{h}\d{s}a\d{n}m\={a}so dak\d{s}i\d{n}enaiti
\d{s}a\d{d}uttare\d{n}a}.

The Br\={a}hma\d{n}as count days starting from the
winter solstice and the period assumed between the two
solstices is 183 days. It is natural to adopt
the equation given above with these conditions which
are part of the old Vedic astronomical tradition.
Use of it in either region does not imply borrowing because
it is the most obvious function to use.

\subsubsection*{The Use of the Water Clock}

The use of the water-clock occurs
in the Atharvaveda 19.53.3 in the expression:$^{30}$

{\it p\={u}r\d{n}a\d{h} kumbho'dhi k\={a}la \={a}hita\d{h}}:
A full vessel is placed upon k\={a}la (time).

The objective of this mantra is to exhort that
``a full vessel be set [up] with reference to 
the [measurement of] time.''

Since the Atharvaveda is prior to the
period of Babylonian astronomy by any account, it shows
that India used water-clocks. Babylonia may have had its
own independent tradition of the use of water-clocks.

\subsubsection*{The Concept of {\it tithi}}

The division of year into equal parts of 30 portions is to
be found in several places in the Vedas and the subsequent
ancillary texts.

In (RV 10.85.5), it is stated that the moon shapes
the year.
In 
Taittir\={\i}ya Br\={a}hma\d{n}a 
the correct technical sense
of tithi is given at
many places. 
For example, in 1.5.10, it is said
that {\it candram\={a} vai pa\~{n}cada\'{s}a\d{h}. e\d{s}a
hi pa\~{n}cada\'{s}y\={a}mapak\d{s}\={\i}yate. 
pa\~{n}cada\'{s}y\={a}m\={a}p\={u}ryate},
``the moon wanes in fifteen, and waxes in fifteen [days].''
In
3.10, 
the fifteen tithis of the waxing moon and 
fifteen tithis of the waning moon are named.

The idea of a tithi is abstract. There are only 27 moonrises
in a month of 29.5 days. To divide it into 30 parts means that
a tithi is smaller than a day. The reason it arose in
India was due to its connection to Soma ritual. 

The number 360 is fundamental to Vedic thought. It represents
the equivalence between time and the subject. In \={A}yurvda,
the number of bones of the developing fetus are taken to be
360.

Since all the six concepts were already in use
in the Sa\d{m}hit\={a}s, in an epoch earlier 
than 1000 BC in the least, they could not have been
learnt by the Indians from the Babylonians who
came to use these concepts after 700 BC.

\section*{Babylonian Observations and Siddh\={a}ntic Astronomy}
Another issue related to the possible connection between
Babylonian and Indian astronomy is whether the excellent
observational tradition of the Babylonians was useful
to the Indians.
Were
ideas at the basis of \={A}ryabha\d{t}a's astronomy
were borrowed from outside or were part of India's
own tradition.
A few years ago,$^{31}$ Abhyankar argues that
``\={A}ryabha\d{t}a's values of {\it bhaga\d{n}as}
were probably derived from the Babylonian planetary
data.''
But Abhyankar makes contradictory assertions in the
paper, suggesting at one place that \={A}ryabha\d{t}a
had his own observations and at another place
that he copied numbers without understanding, making
a huge mistake in the process.

In support of his theory, Abhyankar
claims that \={A}ryabha\d{t}a used
the Babylonian value of 44528
synodic months in 3600 years as
his starting point. But this value
is already a part of the \'{S}atapatha
altar astronomy reconciling lunar and
solar years in a 95-year yuga.
In this ritual, an altar is built to an area
that is taken to represent the nak\d{s}atra
or the lunar year in tithis and the
next design is the same shape but to a larger
area
(solar year in tithis), but since this second
design is too large, the altar construction
continues in a sequence of 95 years.
It appears that satisfactory reconciliation
by adding intercalary months to the
lunar year of 360 tithis amounted to subtracting
a certain number of tithis from the
372 tithis of the solar year, whose most likely
value was 89 tithis in 95 years.$^{32}$

The areas of the altars increase from $7\frac{1}{2}$
to $101\frac{1}{2}$ in the 95 long sequence in
increments of one. The average size of the altar
is therefore
$54\frac{1}{2}$, implying that the average difference
between the lunar and the solar year is taken to
be one unit with
$54\frac{1}{2}$ which is about $6.60$ tithis for
the lunar year of 360 tithis.
This is approximately correct.

Considering a correction of 89 tithis in 95
years,
the corrected length of the
year is $372 - 89/95 = 371.06316$ tithis.
Since each lunation occurs in 30 tithis, the number
of lunations in 3600 years is 44527.579.
In a Mah\={a}yuga, this amounts to
53,433,095. In fact, the number chosen by
\={A}ryabha\d{t}a (row 1 in Table 4)
is closer to this number
rather than the Babylonian number
of 53,433,600.
One may imagine that \={A}ryabha\d{t}a was
creating a system that was an improvement on the
earlier altar astronomy.

Table 4 presents the Babylonian numbers
given by Abhyankar together with
the \={A}ryabha\d{t}a constants
related to the synodic lunar months and the
revolutions of the lunar node, the lunar apogee,
and that of the planets.
It should be noted that
the so-called Babylonian numbers are not actually
from any Babylonian text but were computed by
Abhyankar using the rule of three on various
Babylonian constants.

\vspace{0.2in}

\vspace{0.4in}
{\it Table 4}: Reconstructed Babylonian and \={A}ryabha\d{t}a parameters

\vspace{0.3in}

\begin{tabular}{||r|r|r||} \hline
Type & Babylonian & \={A}rybha\d{t}a \\ \hline
Synodic lunar months & 53,433,600 & 52,433,336 \\ \hline
Lunar node & -232,616 & -232,352 \\ \hline
Lunar apogee & 486,216 & 488,219 \\ \hline
Mercury & 17,937,000 & 17,937,020 \\ \hline
Venus & 7,022,344 & 7,022,388 \\ \hline
Mars & 2,296,900 & 2,296,824 \\ \hline
Jupiter & 364,216 & 364,224 \\ \hline
Saturn & 146,716 & 146,564 \\ \hline
\end{tabular}

\vspace{0.2in}
We see that no numbers match.
How does one then make the case that
\={A}ryabha\d{t}a obtained
his numbers from a Babylonian
text?
Abhyankar says that
these numbers are
different because of his
(\={A}ryabha\d{t}a's)
own
observations
``which are more accurate.''
But if \={A}ryabha\d{t}a had his
own observations, why did he have to
``copy" Babylonian constants, and end up not using them,
anyway?

Certain numbers have great discrepancy,
such as those of the lunar apogee, which
Abhyankar suggests was due to a ``wrong
reading of 6 by 8'' implying--in opposition
to his earlier view in the same paper that \={A}ryabha\d{t}a
also had his own observations--
that \={A}ryabha\d{t}a did not
possess his own data and that he
simply copied numbers from some manual
brought from Babylon!

The \={A}ryabha\d{t}a numbers are also more
accurate that Western numbers as in the work
of Ptolemy.$^{33}$
Given all this,
there is no credible case to accept
the theory of borrowing of these
numbers from Babylon.

Abhyankar further suggests
that \={A}ryabha\d{t}a may
have borrowed from Babylon
the two central
features of his system: (i) the concept of
the Mah\={a}yuga, and (ii) mean
superconjunction of all planets at
some remote epoch in time.
In fact, Abhyankar repeats here
an old theory of Pingree$^{34}$ and
van der Waerden$^{35}$ about a
transmission from Babylon of these
two central ideas.
Here we show that these ideas were already
present in the pre-Siddh\={a}ntic
astronomy and, therefore, a
{\it contrived} connection with
Babylonian tables is unnecessary.

In the altar ritual 
of the Br\={a}hma\d{n}as,$^{36}$ 
equivalences by number connected the altar area to the
length of the year.
The 5-year yuga is described in the
Ved\={a}\.{n}ga Jyoti\d{s}a, where
only the motions of the sun and the moon
are considered.
The \'{S}atapatha Br\={a}hma\d{n}a 
describes the 95-year cycle to harmonize
the solar and the lunar years.
The \'{S}atapatha Br\={a}hma\d{n}a
also describes an asymmetric circuit for
the sun$^{37}$, which the Greeks speak about
only around 400 BC.

Specifically, we find mention of the nominal year of
372 tithis, the nak\d{s}atra year of 324 tithis, and a solar
year of 371 tithis.
The fact that a further correction was required in 95 years
indicates that these figures were in themselves considered to be
approximate.

In the altar ritual, the primal person is made to an area of
$7 \frac{1}{2}$ puru\d{s}as, when a puru\d{s}a is also equated with
360 years leading to another cycle of 2700 years.
This is the 
Saptar\d{s}i cycle which was taken to start and end with a 
superconjunction.

The \'{S}atapatha Br\={a}hma\d{n}a 10.4.2.23-24 describes
that the \d{R}gveda has 432,000 syllables,
the Yajurveda has 288,000 and the
S\={a}maveda has 144,000 syllables.
This indicates that larger yugas in proportion
of 3:2:1 were known at the time of the
conceptualization of the Sa\d{m}hit\={a}s.

Since the nominal size of the \d{R}gveda was considered to be
432,000 syllables (SB 10.4.2.23)
we are led to the theory of a much larger
yuga of that extent in years since the \d{R}gveda
represented the universe symbolically.

Van der Waerden$^{38}$
has speculated that a primitive epicycle theory
was known to the Greeks by the time of Plato.
He suggested such a theory might have been known in the
wider Indo-European world by early first millennium BC.
With new ideas about the pre-history of the
Indo-European world emerging, it is possible to
push this to an earlier millennium.
An old theory may be the source
which led to the development of very different epicycle
models in Greece and India.

The existence of an independent tradition of observation of
planets and a theory thereof as suggested by our analysis of the
\'{S}atapatha Br\={a}hma\d{n}a helps explain the puzzle why the classical
Indian astronomy of the Siddh\={a}nta period uses many constants that
are different from those of the Greeks.

\section*{More on the Great Year}

Since the yuga in the Vedic and the Br\={a}hma\d{n}a periods
is so clearly obtained from an attempt to
harmonize the solar and the lunar years, it
appears that the consideration of the
periods of the planets was the basis of the creation of an
even longer yuga.

There is no reason to assume that the periods
of the five planets were unknown during
the Br\={a}hma\d{n}a age. I have
argued that the astronomical numbers in
the organization of the \d{R}gveda 
indicate with high probability the knowledge
of these periods in the \d{R}gvedic era
itself.$^{39}$ 

Given these periods, and the various yugas related
to the reconciliation of the lunar and the solar years,
we can see how the
least common multiple of these periods will
define a still larger yuga.

The Mah\={a}bh\={a}rata and the Pur\={a}\d{n}as
speak of the kalpa, the day of Brahm\={a},
which is 4,320 million years long.
The night is of equal length, and 360
such days and nights constitute a ``year'' of
Brahm\={a}, and his life is 100 such years
long.
The largest cycle is 311,040,000 million years
long at the end of which the world is
absorbed within Brahman, until another cycle
of creation. 
A return to the initial conditions (implying
a superconjunction) is inherent in such
a conception.
Since the Indians and the Persians were
in continuing cultural contact, it 
is plausible that this was how this old tradition 
became a part of the heritage of the Persians.
It is not surprising then to come across the idea of the World-Year
of 360,000 years in the work of Ab\={u} Ma'shar, who also
mentioned a planetary conjunction in February 3102 BC. 

The theory of the transmission of
the Great Year of 432,000 years,
devised by Berossos, a priest in
a Babylonian temple, to India in about 300 BC,
has also been advanced.
But we see this number being used in
relation to the Great Year in the
\'{S}atapatha Br\={a}hma\d{n}a, a long time
before Berossos.

The idea of superconjunction seems to be at the basis
of the cyclic calendar systems in India.
The  \'{S}atapatha Br\={a}hma\d{n}a speaks of a marriage between the
Seven Sages, the stars of the Ursa Major, and the K\d{r}ttik\={a}s;
this is elaborated in the Pur\={a}\d{n}as where it is stated that the
\d{r}\d{s}is remain for a hundred years in each nak\d{s}atra.
In other words, during the earliest times in India there existed a
centennial calendar with a cycle of 2,700 years.
Called the Saptar\d{s}i calendar,
it is still in use in several parts of India. Its current beginning is taken
to
be 3076 BC.

The usage of this calendar more than 2000 years ago is 
confirmed by the
notices of the Greek historians Pliny and
Arrian who suggest that, during the
Mauryan times, the Indian calendar began in 6676 BC.
It seems quite certain that this was the Saptar\d{s}i calendar with
a beginning which starts 3600 years earlier than
the current Saptar\d{s}i calendar.

The existence of a real cyclic calendar shows that the idea
of superconjunction was a part of the Indic tradition
much before the time of Berossos.
This idea was used elsewhere as well but, given the
paucity of sources, it is not possible to trace a
definite place of origin for it.

\section*{On Observation in Indian Astronomy}

The use of the lunar zodiac creates complicating factors
for observation which were not appreciated by early
historians of Indian astronomy.
Roger Billard's demonstration$^{40}$ of the falsity of 
the 19th century notion that India did not
have observational astronomy has devastating consequences
for the schoolbook histories of early astronomy.
His analysis of the Siddh\={a}ntic and the practical kara\d{n}a
texts 
demonstrated that these texts
provide a set of elements from which the planetary
positions for future times can be computed.
The first step in these computations is the determination of the
mean longitudes which are assumed to be linear functions of time.
Three more functions, the vernal equinox, the lunar node and the
lunar apogee are also defined.

Billard investigated these linear functions for the five planets,
two for the sun (including the vernal equinox) and three for the
moon.
He checked these calculations against the values derived from
modern theory and he found that the texts provide very accurate
values for the epochs when they were written.
Since the Siddh\={a}nta and the kara\d{n}a models are not accurate,
beyond these epochs deviations build up.
In other words, Billard refuted the theory that there was no
tradition of observational astronomy in India.
But Billard's book is not easily available in
India, which is why 
the earlier theory has continued to do rounds
in Indian literature.

\={A}ryabha\d{t}a's constants are more accurate than
the one's available in the West at that time. He
took old Indic notions of the Great Yuga and of
cyclic time (implying superconjunction) and 
created a very original and novel siddh\={a}nta.
He presented  the rotation information of the outer
planets with respect
to the sun -- as was done by the \'{s}\={\i}ghroccas
of Mercury and Venus for the inferior planets --  which means that his system
was partially heliocentric.$^{41}$
Furthermore, he considered the earth to be
rotating on its own axis.
Since we don't see such an advanced system amongst
the Babylonians prior to the time of \={A}ryabha\d{t}a,
it is not reasonable to look outside of
the Indic tradition or \={A}ryabha\d{t}a himself
for the 
data on which these ideas were based. 

The observational protocols used in Indian
astronomy has become an interesting question to be
investigated further.

\section*{Conclusions}

The debate on the relationship between the
astronomical sciences of India and Babylon
became vitiated by the race and colonial theories of
19th century Indologists.
Furthermore, their analysis was done using 
simplistic ideas about cultural interaction.
They took knowledge to flow from one direction to another
without recognizing that the reality was likely to have
been complex and interaction bidirectional.
Considering that a time range of several centuries
was involved and 
interaction through
intermediaries constituted a complex process,
the answer to any question of borrowings
and influence can only be
complicated.

Our review of Indian astronomy shows
that in the period of the early Vedic texts, that
are definitely prior to 1000 BC, the following
facts were known:

\begin{itemize}

\item
Vedic astronomy tracked the motion of the sun and the moon
against the backdrop of the nak\d{s}atras.
The sky was divided into 12 segments (\={A}dityas) and
27 segments (lunar nak\d{s}atras) where the nak\d{s}atra
and deity names were used interchangeably.

\item
Although the names of the solar zodiacal signs (r\={a}\'{s}is) are seen first
in the siddh\={a}ntic texts, we see
they can be derived from the deity names of
the lunar nak\d{s}atra segments. Given that the nak\d{s}atra names
are to be found in the S\d{m}hit\={a}s and the r\={a}\'{s}i
names are not in the Vedic books, one may conclude
that the specific names were chosen sometime in the first
millennium BC, replacing the earlier
\={A}ditya names. But the solar signs were a very early component
of Vedic astronomy, acknowledged in the \d{R}gvedic hymn itself which
speaks of the twelve division of the 360-part wheel of time.

\item
Astrology (Jyot\d{s}a) is a part of the earliest Vedic texts.
Vedic ritual is associated with the time of the day, the nak\d{s}atras,
and the position of the moon.
The year is divided into the deva nak\d{s}atras and yama nak\d{s}atras
and this division is carried down to smaller scales, implying that
certain time durations are more auspicious than others.
The meanings of the nak\d{s}atras (such as
Anur\={a}dh\={a}, ``propitious,'' and Punarvas\={u}, ``giving
wealth'') provide us evidence
that they had an astrological basis.
Vedic nak\d{s}atras are assigned different qualities
and in the marriage ritual the astrologer recommended an
auspicious time.
Taittir\={\i}ya Br\={a}hma\d{n}a 3.1.4 lists
the effects of propitiating the different nak\d{s}atras.

The fundamental notion in Vedic thought is the
equivalence or connection ({\em bandhu})
amongst the adhidaiva ({\em devas} or stars), adhibh\={u}ta
(beings), and adhy\={a}tma (spirit).
These connections, between the astronomical, the terrestrial,
the physiological and the psychological, represent the subtext without
which the Vedas cannot be understood.
These connections are usually stated in terms of vertical
relationships, representing a
recursive system; but they are also described horizontally across
hierarchies where they represent metaphoric or structural
parallels.
Most often, the relationship is defined in terms of numbers or other
characteristics. An example is the 360 bones of the infant---which later
fuse into the 206 bones of the adult---and the 360 days of the year.
Likewise, the tripartite division of the cosmos into earth,
space, and sky is reflected in the tripartite psychological
types.$^{42}$

The bandhu are the rationale for astrology. Indeed, they inform
us that astrology central to the world-view of the Indians.

\item
The use of the tithi system, the division of
the lunar month into 30 parts, is closely connected to Soma
worship, a uniquely Indian ritual.
There is no such ritual connection with the tithis that we know of in
the Babylonian context.

\end{itemize}

The evidence suggests that the Indian ideas of
sacrifice, 12 divisions of the solar year, and the 30 divisions
of the lunar month, and the zodiac reached Babylonia sometime
in the early
first millennium BC.
These new ideas, including the Indian ratio of 3:2 for the
longest to shortest day of the year, 
triggered a new phase of careful observations in
Babylonia which was
to influence astronomy in a fundamental way.

But it is also possible that the Babylonian 
flowering was quite independent based on the
presence of general ideas which were present
in the lands across India to Greece.
In any event, the borrowing was of the most general
ideas, the actual methods were a continuation 
of local tradition.
When we consider the details, we find
the astronomical systems of India and Babylon (and also Greece)
each
have unique features.

Astrology was a part of the ancient world everywhere including
Mesopotamia.
But it appears that the solar zodiacal signs as we know
them originated
in India, where they have a relationship with the deities
of the nak\d{s}atras. They seem 
to have been later adopted in Babylonia in the
middle of the first millennium BC  and subsequently in
Greece.
But this does not mean that the practice of Vedic astrology
was adopted by the Babylonians.

Subsequent to the establishment of the Indo-Greek states on the borders
of India after Alexander, the interaction between Indian and 
Western astrology and astronomy entered a new phase.
Increased political and trade interaction made it possible for
texts to be exchanged.

\subsection*{Acknowledgement}
I am grateful to Narahari Achar and David Frawley for their advice and
criticism.

\newpage
\section*{Notes and References}

\begin{enumerate}

\item
S. Kak, Akhenaten, Surya, and the \d{R}gveda, Louisiana
State University, 2003.\\
http://www.ece.lsu.edu/kak/akhena.pdf

\item
T. Burrow, ``The proto-Indoaryans.''
{\it J. of the Royal Asiatic Society,}
vol. 2, pages 123-140, 1973.
Burrow's idea of proto-Indoaryans is not supported by
current evidence.

\item
S.N. Sen, ``Surveys of Studies in European Languages.''
{\it Indian Journal of History of Science,} vol 20, 49-121, 1985.

\item
E. Leach, ``Aryan invasions over four millennia.''
In {\it Culture through Time, Anthropological
Approaches}, E. Ohnuki-Tierney (ed.), Stanford, 1990, pp. 227-245.
In this Leach, a distinguished anthropologist, 
suggests that racism was behind the idea of the Aryan invasion theory:

\begin{quote}
        Why do serious scholars persist in believing in the
        Aryan invasions?... Why is this sort of thing
        attractive? Who finds it attractive? Why has
        the development of early Sanskrit come to be
        so dogmatically associated with an Aryan
        invasion?...

        Where the Indo-European philologists are concerned,
        the invasion argument is tied in with their assumption
        that if a particular language is identified as having
        been used in a particular locality at a particular
        time, no attention need be paid to what was there
        before; the slate is wiped clean. Obviously, the easiest
        way to imagine this happening in real life is to have
        a military conquest that obliterates the previously
        existing population!

        The details of the theory fit in with this
        racist framework...
        Because of their commitment to a unilineal segmentary
        history of language development that needed to be
        mapped onto the ground, the philologists took it
        for granted that proto-Indo-Iranian was a language
        that had originated outside either India or Iran.
        Hence it followed that the text of the Rig Veda was
        in a language that was actually spoken by those who
        introduced this earliest form of Sanskrit into
        India. From this we derived the myth of the
        Aryan invasions. QED.
\end{quote}

Leach's cry is in the face of the stubborn manner in which some
scholars have persisted with the invasion model although it
flies against archaeological, skeletal, and cultural evidence.
See
Shaffer, J. and D.L. Lichtenstein. ``The concept of cultural
tradition and paleoethnicity in South Asian archaeology.''
In {\it The Indo-Aryans of
South Asia,} G. Erdosy. ed. Berlin: Walter de Gruyter, 1995.
For an overview of this general evidence, see
G. Feuerstein, S. Kak, D. Frawley,
{\it In Search of the Cradle of Civilization.}
Quest Books, Wheaton, 2001;
S. Kak, {\it The Wishing Tree.}
Munshiram Manoharlal, New Delhi, 2001.

\item
D. Pingree, ``History of mathematical astronomy in India.''
In {\em Dictionary of Scientific Biography}, C.S. Gillespie, ed., 533-633.
New York: Charles Scribner's Sons, 1981.
For a critique of such as approach to India, see
N. Dirks, {\it Castes of Mind.}
Princeton University Press, Princeton, 2001.
Says Dirks: 
``Orientalist
generalizations to articulate the justifications for 
permanent colonial rule ... [using] the racialized language
systems and conceits of late nineteenth century imperial
world systems'' led to a major revision in the
Western narratives on Indian science and society.
He adds:
``Colonialism was made possible, and then
sustained and strengthened, as much by cultural 
technologies of rule as it was by the more obvious and brutal
modes of conquest.... Colonialism was itself a cultural project
of control. Colonial knowledge both enabled conquest and was
produced by it; in certain important ways, knowledge was what
colonialism was all about.''

\item
Pingree suggests indirectly ({\it op cit}) that the \d{R}gvedic hymns with the
360 divisions of the year must be dated around
500 BC because this knowledge arose in Babylonia! Elsewhere,
Pingree in his ``Astronomy in India'' (In C. Walker, {\it Astronomy
Before the Telescope.} St. Martin's Press, New York, 1996, page 124)
suggests that the number 360 came from the Babylonian text
{\sc mul.apin} which in its present form is supposed to be no
earlier than 1000 BC.
Pingree's suggestions that Indian text are a mass of 
undifferentiable material is incorrect.
There is much material in
the Indian textual tradition that has had no interpolations, such as
P\={a}\d{n}ini's grammar and the s\={u}ktas of the Sa\d{m}hit\={a}s and
other Vedic texts.

\item 
E. Burgess, (tr.) {\it The S\={u}rya Siddh\={a}nta}.
Delhi: Motilal Banarsidass, 1989 (1860).
For knowledge of planets in India in the 3rd millennium BC.

\item S. Kak, ``Knowledge of planets in the third millennium B.C.''
{\it Quarterly Journal of the Royal Astronomical Society,} 
vol. 37, 709-715, 1996.

\item
D. Pingree, ``The Mesopotamian origin of early Indian
mathematical astronomy.'' {\it Journal of the History of
Astronomy,} vol. 4, 1-12, 1973.
D. Pingree, ``The recovery of early Greek astronomy from
India.'' {\it Journal for the History of Astronomy,} vol 7,
109-123, 1976.

\item
D. Pingree, 
In {\em Dictionary of Scientific Biography}, {\it op cit}.
This goes against linguistic, historical, and geographical evidence
in the Vedic texts and would be rejected by any scholar of Indian
literature. For details of the chronology of Indian texts, see
M. Winternitz, {\it History of Indian Literature.} Orient Books
Reprint, New Delhi, 1972 (1927).
Winternitz believes that the beginning of the texts could not
be pushed later than 2500 BC (see page 310 of volume 1).

\item
S.B. D\={\i}k\d{s}ita, {\it Bh\={a}rat\={\i}ya
Jyoti\'{s}\'{s}\={a}stra.} Hindi Samiti, Lakhanau, 1963 (1896).

\item
D. Pingree, ``Astronomy and astrology in India and Iran,''
{\it Isis,} vol 54, 229-246, 1963.

\item
D. Pingree, ``The Mesopotamian origin of early
Indian mathematical astronomy,'' 1973, {\it op cit}, page 10.

\item
See, for example,
N. Achar,  ``On the Vedic origin of the ancient
mathematical astronomy of India.''
{\it Journal of Studies on Ancient India},
vol 1, 95-108, 1998.

\item Roger Billard, {\it L'astronomie Indienne.}
Paris: Publications de l'ecole francaise d'extreme-orient, 1971.

\item
B.L. van der Waerden, ``Two treatises on Indian
astronomy.'' {\it Journal for History of Astronomy}
vol 11: 50-58, 1980.

\item S. Kak, {\it The Astronomical Code of the \d{R}gveda.}
New Delhi: Munshiram Manoharlal, 2000;\\
S. Kak, ``Birth and early development of Indian
astronomy.'' In {\it Astronomy Across Cultures: The History
of Non-Western Astronomy}, H. Selin (ed.), pp. 303-340.
Dordrecht: Kluwer Academic Publishers, 2000.

\item
O. Neugebauer, {\it A History of Ancient Mathematical Astronomy.}
Springer-Verlag, New York, 1975;
A. Aaboe, {\it Episodes from the Early History of Astronomy.}
Springer, New York, 2001.

\item
G.C. Pande (ed.), {\it The Dawn of Indian Civilization.}
       Centre for Studies in Civilizations, New Delhi, 2000.

\item
N. Achar, ``On the identification of the Vedic nak\d{s}atras,''
In {\it Contemporary Views on Indian Civilization},
Proceedings of the WAVES
Conference, held in Hoboken NJ, 2000. (ed) BhuDev Sharma,  pp 209-229.

\item 
N. Achar, ``In search for nak\d{s}atra in \d{R}gveda,'' 
In {\it Contemporary Views on Indian Civilization},
Proceedings of the WAVES
Conference, held in Hoboken NJ, 2000. (ed) BhuDev Sharma,  pp 200-208.

\item
S. Kak, {\it The A\'{s}vamedha: The Rite and Its Logic.}
Motilal Banarsidass, Delhi, 2002.

\item 
See Kak, {\it The Astronomical Code,} {\it op cit.}

\item
K.D. Abhyankar, ``Uttar\={a}ya\d{n}a.'' In
{\it Scientific Heritage of India,} B.V. Subbarayappa and
S.R.N. Murthy (eds.). The Mythic Society, Bangalore, 1988.

\item  T.S. Kuppanna Sastry, {\em Ved\={a}\.{n}ga Jyoti\d{s}a of Lagadha.}
New Delhi: Indian National Science Academy, 1985.
The 1800 BC date for it is argued in 
B.N. Narahari Achar, ``A case for revising the date of 
Ved\={a}\.{n}ga Jyoti\d{s}a,''
{\it Indian Journal of History of Science,} vol. 35, pp. 173-183, 2000.

\item 
Aaboe, {\it op cit.}

\item
O. Neugebauer, {\it op cit.}

\item 
Pingree, 1973, {\it op cit}, page 9.

\item
B.N. Achar, 1998, {\it op cit.}

\item
B.N. Achar, {\it op cit.}

\item 
K.D. Abhyankar, ``Babylonian source of \={A}ryabha\d{t}a's
planetary constants.''
{\it Indian Journal of History of Science,}
vol 35, 185-188, 2000.

\item S. Kak, {\it The Astronomical Code of the \d{R}gveda,}
page 87.

\item K.S. Shukla and K.V. Sarma, {\em \={A}ryabha\d{t}\={\i}ya of
\={A}ryabha\d{t}a.}
New Delhi: Indian National Science Academy, 1976.

\item D. Pingree,  1963, {\it op cit.}

\item B.L. van der Waerden, ``The great year in Greek, Persian
and Hindu astronomy.'' {\it Archive for History of Exact 
Sciences,} vol 18, 359-384, 1978.

\item S. Kak, {\it The Astronomical Code of the \d{R}gveda,}
or
S. Kak, ``Birth and early development of Indian
astronomy,'' {\it op cit.}

\item S. Kak, ``The sun's orbit in the Br\={a}hma\d{n}as.''
{\it Indian Journal of History of Science},
vol 33, 175-191, 1998.

\item  B.L. van der Waerden, ``The earliest form of the epicycle
theory.'' {\it Journal for the History of Astronomy} vol 5,
175-185, 1974.

\item S. Kak, {\it The Astronomical Code of the \d{R}gveda,}
{\it op cit.}


\item Roger Billard, {\it op cit.}

\item H. Thurston, {\it Early Astronomy.}
New York: Springer-Verlag, 1994, page 188.

\item
See Kak, {\it Astronomical Code}, {\it op cit},
or S. Kak, {\it The Wishing Tree,} {\it op cit.}

\end{enumerate}

\vspace{0.2in}
\noindent
\copyright Subhash Kak
\end{document}